# Computationally efficient methods for fitting mixed models to electronic health records data

KM Rhodes[1]; RM Turner[1,2]; RA Payne[3]; IR White[1,2]

[1]MRC Biostatistics Unit, University of Cambridge, Cambridge Institute of Public Health, Cambridge, UK
[2]MRC Clinical Trials Unit at University College London, Institute of Clinical Trials & Methodology, London, UK
[3]Centre for Academic Primary Care, School of Social and Community Medicine, University of Bristol, Bristol, UK

Motivated by two case studies using primary care records from the Clinical Practice Research Datalink, we describe statistical methods that facilitate the analysis of tall data, with very large numbers of observations. Our focus is on investigating the association between patient characteristics and an outcome of interest, while allowing for variation among general practices. We explore ways to fit mixed effects models to tall data, including predictors of interest and confounding factors as covariates, and including random intercepts to allow for heterogeneity in outcome among practices. We introduce: (1) weighted regression and (2) meta-analysis of estimated regression coefficients from each practice. Both methods reduce the size of the dataset, thus decreasing the time required for statistical analysis. We compare the methods to an existing subsampling approach. All methods give similar point estimates, and weighted regression and meta-analysis give similar standard errors for point estimates to analysis of the entire dataset, but the subsampling method gives larger standard errors. Where all data are discrete, weighted regression is equivalent to fitting the mixed model to the entire dataset. In the presence of a continuous covariate, meta-analysis is useful. Both methods are easy to implement in standard statistical software.

# 1. Introduction

Routinely collected datasets including electronic health records and other administrative datasets are becoming increasing widely used in healthcare research. These data sources can offer a number of advantages over traditional designed data sources such as randomised trials and surveys, for example speed of access, richness of data recording, extended longitudinal measurements and low cost. Traditionally designed studies are often limited in size and scope; they may exclude patients with comorbidities and the elderly, and patients may decline to participate. Using routine data means that research can be done on a much wider population and therefore with much greater precision, often using hundreds of thousands of patient records to answer questions about what is happening in the "real world".

In the UK, development of electronic health records databases has been aided by the state-funded National Health Service (NHS); virtually all UK residents are registered with a primary care based general practitioner (family doctor) who provides care to all ages and acts as a gatekeeper to specialist secondary care services. Importantly, there is near universal adoption of clinical computer systems by general practices. The Clinical Practice Research Datalink (CPRD) is a government supported initiative to provide researchers with access to NHS data in a secure and ethical way. The CPRD database represents one of the largest longitudinal primary care datasets in the world[1], collating routinely collected anonymised electronic health records data from consenting general practices on a monthly basis. As of March 2015, CPRD included over 11.3 million patients from 674 general practices, representing approximately 7% of the UK population.[1]

Electronic health records such as those provided by CPRD are important in shaping public health policy. They are used to identify patients at risk of diseases; monitor the safety of medicines and vaccinations; and understand the effectiveness of treatments in different groups of patients. For this reason it is essential that studies using CPRD data are carried out to a high standard. In practice, however, routine datasets are subject to a number of challenges because they have not been collected with a specific research question in mind. These challenges include heterogeneity among outcomes, missing data and confounding; if left unaddressed, these could result in bias and incorrect inferences. Methods such as generalized linear mixed models, multiple imputation[2] and propensity score adjustment[3] can be used to address these limitations. However, computational constraints can limit their applicability to datasets comprising very large numbers of observations. Datasets extracted for the purposes of research are usually of a size that can be stored in a standard desktop computer's memory, but it can take a number of hours to fit standard statistical models to data containing a very large number of observations, also known as tall data. This is less of an issue when one wants to fit a single statistical model, but in practice researchers using routinely collected data often fit several models, for example to compare different statistical models, investigate effect modification and conduct sensitivity analyses. As datasets are increasing in size and statistical models are becoming more complex, there is a need for computationally efficient methods for analysing tall health data.

Recently, new statistical and/or computational methodologies have been proposed that scale problems to a reasonable size[4]. These include the "divide and conquer" approach, where the data are divided into subsamples, the subsamples are analysed in parallel and the results are then combined across subsamples.[5,6] Other methods include the use of experimental design techniques to extract a representative subsample[7] and the "bags of little bootstrap" approach[8], where results from bootstrapping a number of subsamples are averaged to give the effect estimate of interest. These methods would be difficult for applied researchers with only a basic understanding of statistics and programming to understand and implement.

Our focus is on fitting statistical regression models to routine clinical data, where the computing time required to fit such models is long due to a large number of observations and heterogeneity in outcome among very many healthcare institutions (such as general practices). We explore the use of weighted regression and meta-analysis techniques; these statistical methods are likely to be familiar to applied researchers in healthcare research, though not for the purpose of analysing tall routine datasets. We compare these methods to a recently proposed subsampling approach.[7] The methods are applied to electronic health records data from two CPRD studies, where the interest lies in investigating the association between some patient characteristics and a health-related outcome. We describe the advantages and disadvantages of subsampling, weighted regression and meta-analysis approaches, and identify the settings in which these methods would be most useful. $R$[9] computing code for all methods is available in Supporting Information.

The rest of the paper is set out as follows. In section 2 we introduce two case studies using electronic health records data from CPRD for research. In section 3, methods for tall data analysis are described, including an existing subsampling approach, weighted regression and meta-analysis approaches. The use of these methods is illustrated in section 4 through application to the case study examples described in section 2. We conclude with a discussion in section 5.

**2. Case studies using electronic health records data**

**Case study 1**
Electronic health records data from CPRD were used to investigate the impact of the UK primary care payment-for-performance system (Quality and Outcomes Framework, QOF) on the detection and treatment of cardiovascular risk factors in patients with severe mental illness.[10] This retrospective open cohort study included 427,190 patients of at least 35 years of age (67,239 with severe mental illness (SMI); 359,951 without SMI). In the original study, six different binary outcomes were considered, but in this paper we focus on recording of elevated cholesterol, defined as a binary indicator of first recording of serum cholesterol ≥5mmol/L. The dataset used for statistical analysis comprises 2,116,948 observations recorded by 674 general practices in the UK between 1996 and 2014. There are two interventions of interest: the first was a QOF incentive in 2004 for annual review of physical health in patients with SMI, and the second was a revised incentive in 2011 specific to cardiovascular review.

Our analyses were carried out as pre-specified in the original study protocol before data access. We used an interrupted time series analysis, as illustrated in Figure 1. The targets of inference are the changes in intercept and slope in years 2004 and 2011. We chose to use a logistic regression model, for its preferable interpretation as modelling log odds. In the logistic regression model we analysed the binary outcomes $y$, allowing separate intercepts for the cases with SMI and controls without SMI.

For observations indexed $i$ within practices $j$, we used the following linear predictor:

$$\text{logit}(E(y|b)) = \beta_0 + b_j + \beta_1 \times (\text{year}_{ij}-1994) + \beta_2 \times I(\text{year}_{ij} \geq 2004) + \beta_3 \times I(\text{year}_{ij} \geq 2011) + \beta_4 \times I(\text{year}_{ij} \geq 2004) \times (\text{year}_{ij}-2004) + \beta_5 \times I(\text{year}_{ij} \geq 2011) \times (\text{year}_{ij}-2011) + \beta_6 \times \text{SMI}_{ij} + \beta_7 \times \text{SMI}_{ij} \times (\text{year}_{ij}-1994) + \beta_8 \times \text{SMI}_{ij} \times I(\text{year}_{ij} \geq 2004) + \beta_9 \times \text{SMI}_{ij} \times I(\text{year}_{ij} \geq 2011) + \beta_{10} \times \text{SMI}_{ij} \times I(\text{year}_{ij} \geq 2004) \times (\text{year}_{ij}-2004) + \beta_{11} \times \text{SMI}_{ij} \times I(\text{year}_{ij} \geq 2011) \times (\text{year}_{ij}-2011) + \beta_{12} \times \text{age}_{ij} + \beta_{10} \times \text{gender}_{ij,}$$

where the main effects are year, the year of the data recording; SMI, an indicator of SMI; $I(\text{year}_{ij} \geq 2004)$, an indicator for intervention 1; $I(\text{year}_{ij} \geq 2011)$, an indicator for intervention 2.

The parameters of interest are: $\beta_8$, the change in intercept at year 2004 for the SMI vs. non-SMI group; $\beta_9$, the change in intercept at year 2011 for the SMI vs. non-SMI group; $\beta_{10}$, the change in slope after year 2004 for the SMI vs. non-SMI group; $\beta_{11}$, the change in slope after year 2011 for the SMI vs. non-SMI group. Also of interest is the heterogeneity among general practices in the pre-intervention intercept $\beta_0$, since it was hypothesised that baseline data for practices would show important variation. To allow for heterogeneity in outcome among 674 general practices, we included random intercepts $b=(b_1,...,b_{674})$ in the model. Adjustment was made for age (as a continuous variable) and gender as confounding factors in the model.

**Case study 2**

A separate subset of CPRD data was used to investigate the association between multi-morbidity (the co-occurrence of two or more health conditions) and health service utilisation in a cohort of 403,985 adult patients from 404 general practices in England followed for a period of four years, starting in 2012.[11] Outcomes of interest were rates of primary care consultations, prescriptions of medications and hospital admissions. Here we analyse rates of consultations among 349,785 adult patients from 353 general patients. We removed data from 51 practices with zero recordings of consultations because we believe that consultation data are missing for these practices.

We used a negative binomial regression model to model the number of primary care consultations, and included three covariates: age (as a continuous variable), gender and the number of morbidities, categorised into groups of low (0-1), moderate (2-3), high (4-5), very high (6+). An offset variable was used to define the exposure period. It has been shown that there are large variations in the recording of data among practices[1]. Variation between practices was therefore of interest and accounted for by the inclusion of a random intercept for each of 353 general practices in the model.

**3. Methods**

The problem common to our two case studies was the length of time required to fit non-linear mixed effects regression models to data comprising hundreds of thousands of observations nested within hundreds of general practices: 11 hours in case study 1 and 1.7 hours in case study 2, using *R* (version 3.3.1)[9] on a computer with a 64-bit Windows operating system, Intel Core i7 processor, 3.4GHz speed and 16 GB installed memory (RAM).

In this paper we reproduce analyses that have been carried out in two applied research studies using electronic health records data from the CPRD. Both studies used a generalised linear mixed model[12] in the statistical analysis. Here we describe the general form of this model.

Observations on the $i$th of $N$ units consist of an outcome $y_i$ and vectors $\mathbf{x}_i$ and $\mathbf{z}_i$ of explanatory variables associated with the fixed and random effects. Units in health records data are typically clustered, for example patients may be grouped within general practices located in regions, and there can be multiple measurements on the same patient. We suppose that, given a $q$-dimensional vector $\mathbf{b}$ of random effects, the outcomes $y_i$ are conditionally independent. The conditional mean of $y_i$ is related to the linear predictor through a link function $g$:

$$E(\mathbf{y}|\mathbf{b}) = g^{-1}(\mathbf{X}\boldsymbol{\beta} + \mathbf{Z}\mathbf{b}), \qquad (1)$$

where *y* is the vector of outcomes $(y_1,...,y_n)^T$, *X* and *Z* are the design matrices with rows $\mathbf{x}_i^T$ and $\mathbf{z}_i^T$, and *β* is a *p*-dimensional vector of fixed effects.

We assume that *b* has a multivariate normal distribution with mean **0** and covariance matrix Σ.

Generalised linear mixed effects models have become relatively straightforward to fit with the use of readily available computing code for standard statistical regression software, such as the *glmer* function for *R*[9] in library *lme4*[13], and the *melogit* and *menbreg* commands for *Stata*[14]. However, standard regression software has not been developed for the purpose of analysing tall data.

**Subsampling approach**

Instead of analysing all the data, Drovandi *et al.*[7] proposed a subsampling algorithm based on using experimental design techniques to extract a subsample that is representative of the entire dataset.

This approach involves an initial learning phase, in which we learn about our model parameters by fitting the mixed effects regression model to a sample of the entire dataset. One difficulty in using design of experiments to construct a subsample is the need to specify an initial subsample. The choice of initial sample influences how many iterations of the algorithm are required in order to reach a subsample that is sufficient to precisely estimate the parameters of the regression model. For our approach, we took a random sample of 10,000 observations as our initial sample in a similar way to Drovandi et al. Alternatively, the initial sample could be empirically based, for example by ensuring that the difference in the mean outcome in the initial and full sample is small. In this paper, the size of the initial subsample is chosen such that the proposed statistical model can be fitted in reasonable time, and the sample is large enough to allow model parameters to be reasonably well estimated.

Each combination of covariates **x** to be included in the model is referred to as a "design" in the subsampling algorithm described by Drovandi et al.[7] In our implementation of the subsampling algorithm, we assume all combinations of covariates **x** that are present in the entire dataset are available for selection to the subsample. At each value of **x** we evaluate the following utility function to quantify how the precision of the estimated regression coefficients $\hat{\beta}$ would change if we added an extra data point with covariate values **x**:

$$U(\mathbf{x}) = \det(I_0(\beta) + I_X(\beta))$$

where $I_0(\beta)$ is the expected Fisher information matrix for observations in the current subsample and $I_X(\beta)$ is the expected Fisher information matrix for a unique observation with covariate values **x**. We evaluate each Fisher information matrix, $I_0(\boldsymbol{\beta})$ and $I_X(\boldsymbol{\beta})$, at the regression coefficients $\boldsymbol{\beta} = \hat{\boldsymbol{\beta}}$ estimated using the current subsample.

A difficulty in using experimental design techniques is the need to evaluate a utility function, which is typically difficult to compute and may require numerical approximation. When calculating each expected Fisher information matrix, we treat the random effects *b* as nuisance parameters and fix these to zero in our implementation of the subsampling approach. This removes the integrals with respect to *b* in the expected Fisher information and simplifies the computation. We derive the expected Fisher information for the logistic regression and negative binomial models in Supporting Information S1.

We want to find the optimal covariate combination **x**\* that gives the maximum value of *U(***x***)*. For our purpose of parameter estimation, the expected Fisher information matrix is a sensible utility function because maximising the utility function corresponds to minimising the variance of the estimated regression coefficients $\hat{\boldsymbol{\beta}}$ and increasing the amount of information. Since it is easier to work with a scalar than a matrix, we took the determinant of the expected Fisher information in the utility function as an estimate of overall precision, which accounts for correlations among pairs of regression coefficients.

The optimal covariate combination **x**\* is chosen as the covariate combination **x** with the maximum utility *U(***x***)*. We add these data to our subsample and repeat the process as follows:

1. Fit the mixed effects regression model to the current subsample
2. Evaluate the utility function *U(***x***)* at each unique covariate combination **x** remaining in the original dataset.
3. Find the covariate combination/s **x**\*= **x** that maximises *U(***x***)*.
4. Extract observations with covariate combination/s **x**\* remaining in the original dataset.
5. Add these data to the subsample and return to step 1.

This process continues until we reach a desired subsample size or until we collect sufficient information required to answer the research question, e.g. when the standard errors of parameter estimates are reasonably small and there is little change in the estimated parameters with increasing iterations of the algorithm. In the applications to the case study examples, we chose to stop the process at reaching a sample size that was "fair" to this method, e.g. to allow a reasonable number of iterations of the algorithm, but choice of sample size may be based on practical considerations.

In their subsampling algorithm, Drovandi et al.[7] consider all possible combinations of covariate values **x** that are available for inclusion in the subsample, irrespective of whether they were present in the full dataset or not. This helps to identify where information is lacking in the dataset. If the optimal covariate combination is not present in the dataset, the combination present in the dataset that is closest in Euclidean distance to the optimal covariate combination is chosen. In contrast, our approach is to only consider selecting covariate combinations that are present in the entire dataset, since there are likely to be many possible covariate combinations in big datasets and it would be impractical to evaluate the utility function for each of these covariate combinations.

In some datasets the number of covariate combinations present may be small relative to the size of the dataset, and therefore the number of observations per covariate combination is large. This means that at each stage of the subsampling procedure, we add a large number of observations to our current subsample and consequently only a small proportion of covariate combinations are sampled before the desired subsample size is reached. In this situation, we suggest modifying the above subsampling approach to extract a random sample of values with the chosen covariate combination **x** rather than all observations with the chosen covariate combination **x** at each stage of the subsampling procedure. Choice of the proportion of observations to extract from the entire dataset will depend on the number of covariate combinations in the dataset; if this is very small relative to the size of the dataset, then we might want to limit the proportion of observations that we extract with the chosen covariate combination, in order to allow time for other covariate combinations to be potentially selected for inclusion in the subsample.

**Weighted regression**

The idea of weighted regression is to collapse the dataset such that it contains only unique observations, i.e. where no two observations are equivalent in all outcome and covariate values, and a variable indicating the number of times each observation occurs in the full sample.

The first step in this method is to create a new variable to weight our data based on the frequency of observations with the same values of outcome, covariates and any nesting variable (in our case studies general practice). Then we collapse the dataset to a smaller dataset of values of outcome, covariates, nesting variables, and weight. In other words, we remove replicated observations with the same values of outcome, covariates and nesting variables.

Where the outcome of interest is a rate as in case study 2, we also need to weight and collapse the data by the offset variable that is used to denote the exposure period. In the presence of a continuously measured covariate, we categorize the covariate values into groups, for example by inspecting the entire dataset to find sensible cut-off values or quartiles, and replace the values of that covariate by the mean value for observations with the same recorded outcome, covariate combination (including category of continuous covariate) and values of nesting variables. A larger number of groups would better reflect the variation in the continuous covariate values, but the number of groups should be small enough for the original dataset to be collapsed to a sufficiently smaller size for analysis. By assigning each observation the mean value of the continuous covariate for its group defined by covariate combination, outcome and nesting variables, the measurement error introduced by categorizing continuous covariates is a Berkson error (uncorrelated with the covariate values used in the weighted regression analysis) and hence only a very minor source of bias.[15]

In the statistical analysis we use a weighted generalised linear mixed effects model, This method is easy to implement, by specifying the argument *weight* in the function *glmer*[13] for *R*.[9] In a dataset comprising only discrete outcome and covariate data, fitting the weighted model to the collapsed dataset is equivalent to fitting the standard model to the entire untransformed dataset.

**Meta-analysis (Divide and recombine)**

In applications where the data are too large to analyse all at once, it has been proposed to use divide and recombine approaches, in which we divide the data into a number of smaller samples, analyse these subsamples individually, and then combine the results of these individual analyses.[4,5]

Our aim is to estimate the regression coefficients $\beta$ in a multi-level model, in order to determine the association between $p$ independent variables and an outcome of interest. We divide the data into top level units according to a variable $\mathbf{z}$ and fit the regression model (equation (1)) to the data from each subgroup (conditional variable division). Fitting the regression model to data from each subgroup provides estimates $\hat{\beta}_{kj}$ of the regression coefficients $\beta_{kj}$ ($k=0,\ldots,p$) for each subgroup $j$ of $\mathbf{z}$.

We combine the regression coefficient estimates across subgroups using two meta-analysis techniques: univariate meta-analysis and multivariate meta-analysis.

In separate univariate meta-analyses of estimated regression coefficients, we assume that the estimated regression coefficients $\hat{\beta}_{kj}$ from each subgroup $j$ have a normal random-effects distribution:

$$\hat{\beta}_{kj} \sim N(\beta_{kj}, s_{kj}^2)$$
$$\beta_{kj} \sim N(\beta_k, \tau_k^2),$$

where $\beta_{kj}$ is the true regression coefficient for each subgroup, $\beta_k$ is the combined regression coefficient, $s_{kj}^2$ is the estimated within-subgroup variance for the regression coefficient (assumed known) and $\tau_k^2$ is the between-subgroup variance. We estimate $\tau_k^2$ for each regression coefficient $k$ that is assumed to vary across subgroups in equation (1), and set $\tau_k^2$ to be zero for fixed regression coefficients.

Regression coefficients estimated from the same model are correlated and by meta-analysing each regression coefficient separately this correlation is ignored. This could lead to overestimated variances of the combined regression coefficients $\beta_k$ and biased estimates. Multivariate meta-analysis provides a solution to this problem, by summarizing all regression coefficient estimates simultaneously.

For each subgroup *j*, we have *p+1* estimated regression coefficients denoted as $\hat{\beta}_j = (\hat{\beta}_{0j},\ldots,\hat{\beta}_p)'$. In a multivariate meta-analysis, we assume these estimates to have a multivariate normal (MVN) distribution:

$$\hat{\beta}_j \sim MVN(\beta_j, S_j)$$
$$\beta_j \sim MVN(\beta, \Sigma),$$

where $\beta_j$ is the vector of true regression coefficients for subgroup *j*, $\beta$ is the regression coefficient vector combined across subgroups, $S_j$ is the estimated within-subgroup covariance matrix of $\hat{\beta}_j$ (assumed known) containing the *p* variances of the regression coefficients in the diagonal entries and their covariances in the off diagonal entries, and $\Sigma$ is the between-subgroup covariance matrix containing the *p* between-subgroup variances of the regression coefficients in the diagonal entries: $\tau_0^2, \ldots, \tau_p^2$, and their between-subgroup covariances in the off diagonal entries. If all entries of $\Sigma$ are set to zero, then the meta-analysis model reduces to a common effect model, which is similar to reducing the generalized linear mixed model in (1) to a standard generalized linear model. We note that this common effect approach is equivalent to the divide and conquer strategy proposed earlier by Lin and Xi.[6]

In large datasets, there is often scope to examine subgroup-level effects, for example in some clinical datasets potential effect modifiers at the practice level may include: practice size, an index of multiple deprivation, and the number of patients per general practitioner. Subgroup-level covariates $x_j$ can be incorporated in the second stage of the univariate meta-analysis model for the estimated intercept[16]:

$$\hat{\beta}_{0j} \sim N(\beta_{0j}, s_{0j}^2)$$
$$\beta_{0j} \sim N(\beta_0 + \lambda x_j, \tau_0^2).$$

Likewise, in a multivariate meta-analysis we can model the intercept estimates $\beta_{0j}$ in terms of subgroup-level covariates.

In our applications to CPRD data, we divided the data into subgroups by general practice, analysed the data from each practice separately and combined the results from each practice using meta-analysis techniques. The regression models originally fitted to each dataset assumed a random practice-level intercept and fixed covariate effects. For this reason, we estimated between-practice variance in the univariate meta-analysis of intercept estimates only and fixed the between-practice variance to zero for all other univariate meta-analyses of regression coefficients, $k=1,\ldots,p$. Similarly, in the multivariate meta-analysis we only estimated the first entry in $\Sigma$ corresponding to the between-practice variance in intercept and set all remaining entries of to zero.

We implemented univariate and multivariate meta-analyses using the *metafor* package[17] and *mvmeta* package[18] or *R*, respectively. A number of methods are available for estimating the meta-analysis models, including restricted maximum likelihood estimation (REML) and method-of-moments approaches. These methods primarily differ in the estimation of the between-subgroup covariance matrix $\Sigma$. In this paper we present results using REML, which is commonly used for mixed effects models because it yields unbiased estimates of variance and covariance parameters. The difficulty with likelihood-based methods is that they become computationally intensive and time consuming as the number of subgroups and regression coefficients increases. Where the number of regression coefficient estimates per subgroup is large such that computing time for estimation of the multivariate meta-analysis model is very slow, one may prefer to use method-of-moments estimation, which requires no numerical maximization or iteration and is consequently fast to implement.

## 4. Applications to case study examples

In this section we apply the above methods to the two case study examples (Section 2) to demonstrate the use of each method for fitting mixed regression models to tall routine datasets.

**Case study 1**

Results from fitting the logistic regression model to the entire dataset are displayed graphically in Figure 1 and we report results for the parameters of interest numerically in Table 1. We report marginal effects for variables of interest, averaged over all observations in the sample. Marginal effects are calculated by subtracting the conditional predicted probability of the outcome for a non-SMI patient when all covariates are fixed from the same conditional predicted probability, with the indicator of an SMI patient set to 1. Following intervention 1 in 2004, there is evidence of an immediate increase (i.e. an intercept change) in the recording of serum cholesterol ≥5.0mmol/L for the SMI patients, compared to the non-SMI patients: risk difference 0.04 (95% confidence interval (CI): 0.001 to 0.08). This appears to be sustained over time: the average risk difference for a SMI patient vs. non-SMI patient is 0.003 (95% CI: 0.001 to 0.005) in year 2004 and similar at 0.002 (95% CI: <0.001 to 0.004) in year 2010. Intervention 2 in 2011 is associated with a further immediate increase in recognition of cases of serum cholesterol ≥5.0mmol/L (risk difference 0.09, 95% CI: 0.01 to 0.17), but this is not sustained over time (risk difference in year 2011: 0.08, 95% CI: 0.04 to 0.12; risk difference in year 2014: 0.008, 95% CI: 0.004 to 0.01).

Table 1 also reports the results based on using the alternative methods described in section 3. We progress through the subsampling algorithm until at least 200,000 observations (approximately 10% of the size of the entire dataset) have been sampled. At each iteration of the algorithm, we assume that only the observed combinations of the covariate levels shown in Table 2 are available for selection, resulting in 532 covariate combinations **x** at the first iteration of the algorithm. We note that this approach reduces computing time by 39%, compared with the approach which considers all 2128 possible covariate combinations available for selection at each stage of the algorithm (results not shown), as proposed by Drovandi et al.[7]

Figure 2 shows the 95% confidence intervals for all regression coefficient estimates of interest at each iteration of the subsampling algorithm. As expected, the confidence intervals tend to narrow with each additional iteration, as more data are included in the analysis. In total we selected 93 covariate combinations **x** for inclusion in our subsample; Table 3 shows the first ten values selected for each covariate. The number of observations in the entire dataset per covariate combination ranges from 256 to 19,370 with median 2054 (IQR: 909 to 5823). The number of observations extracted at each iteration of the subsampling algorithm ranges from 252 to 17,791 with median 1488 (IQR: 505 to 2233). After reaching 200,000 observations, there are noticeable changes in parameter estimates with each additional iteration of the subsampling algorithm (Figure 2). This indicates that we stopped the algorithm too early at reaching 200,000 observations. It is only after reaching iteration 361 of the subsampling algorithm that there is little change in parameter estimates with increasing iterations. This corresponds to analysing 878,490 observations (42% of the entire dataset). At this point, all 95% confidence intervals include the parameter estimated by analysing the entire dataset.

In this dataset, the number of observations per practice with each combination of covariate values **x** and outcome ranges from 1 to 133 with median 3 (inter-quartile range (IQR) 1 to 19). In the weighted regression approach, removing replicated covariate combinations and corresponding outcomes within each practice reduces the size of the dataset by 87% to 273,685 observations. The results are almost the same as those obtained through fitting the standard model to all observations in the entire dataset, but are obtained in an average time of 4.8 minutes rather than 11 hours on our standard desktop computer (Table 1).

Meta-analysis is the fastest approach to implement, taking just 68 seconds. In this example, correlation between the covariate time and the intercept, and between the indicator of intervention 1 and the intercept, causes the univariate meta-analysis of intercept estimates to yield a higher between-practice variance $\tau^2$ estimate of 2.68, compared with the standard approach of analysing the entire dataset ($\tau^2$=0.05). The multivariate meta-analysis approach incorporates the correlation between main effects and hence yields the same estimate of 0.05 for $\tau^2$ as the conventional analysis of the entire dataset.

The estimated risk differences obtained through conventional analysis of the entire dataset are almost identical to those derived using weighted regression, and results based on subsampling and meta-analysis techniques are similar (Table 1). The standard errors under the subsampling approaches are reasonably small compared with those derived from analyses of the entire dataset.

**Case study 2**

Table 4 displays the results for fitting the negative binomial model to the entire dataset. During the four year follow-up period, patients had a median of 14 primary care consultations (IQR 4 to 32). After adjusting for age and gender, those with a very high (≥6) number of clinical conditions have, on average, higher rates of primary care consultations; rate ratio 3.94 (95% CI: 3.86 to 4.01), compared with the reference group of participants with at most one clinical condition (Table 4). Rates of primary care consultations are also higher, on average, for patients with 4 or 5 clinical conditions (rate ratio 2.95, 95% CI: 2.92 to 2.97), and for patients with 2 or 3 clinical conditions (rate ratio 2.11, 95% CI: 2.10 to 2.14).

Also displayed in Table 4 are results from the different methods described in section 3. At each iteration of the subsampling algorithm, we assume that only the observed combinations of the covariate levels (formed by inspecting the entire dataset) shown in Table 2 are available for selection, resulting in 72 out of a possible 144 covariate combinations **x**. We reach a desired sample size of 35,000 in 39 iterations of the algorithm and select 39 covariate combinations to add to our initial subsample. The number of observations per covariate combination ranges from 2 to 26,497 with median 1523 (IQR: 167 to 3037). The subsampling algorithm tends to select the less frequent covariate combinations (Table 3); the number of observations extracted at each iteration of the subsampling algorithm ranges from 2 to 3007 with median 408 (IQR: 152 to 1015). Results show some discrepancies to those obtained through fitting the regression model to the entire dataset. In particular the estimate of the rate ratio representing the influence of gender (male vs female) on the outcome is 0.87 (95% CI 0.85 to 0.89) based on subsampling and 0.68 (95% CI 0.67 to 0.68) based on the entire dataset. As in case study 1, this gives some indication that we may have stopped the subsampling algorithm too early at 35,000 iterations.

The number of observations per practice with each combination of covariate values, offset variable (exposure time) and outcome ranges from 1 to 275 with median 1 (IQR 1 to 2). We collapse the dataset by groups defined by covariate combination, offset variable, outcome and practice, reducing the size of the dataset to 233,611 observations (67% of the entire dataset). After inputting the mean age for each group, a weighted regression of these data yields approximately identical results to the standard approach of fitting the regression model to all observations in the entire dataset.

Using meta-analysis to combine the estimated regression coefficients across practices leads to very close estimates for the regression parameters of interest, compared to the conventional approach of analysing the entire dataset.

Overall, point estimates for rate ratios derived from the conventional analysis of the entire dataset are close to those derived through alternative approaches using weighted regression and meta-analysis techniques (Table 4). Results from the subsampling approach lead to very similar conclusions of associations between covariates and the outcome of interest, although there are some discrepancies in parameter estimates. The standard errors under the subsampling approaches are reasonably small compared with those obtained through analysing the entire dataset. The directions of the effect estimates are the same across methods, for all rate ratios.

## 5. Discussion

Fitting mixed effects regression models has become relatively straightforward with the use of readily available computing code for standard statistical software such as *R* or *Stata*. However, it is time consuming and impractical to fit such models to tall data comprising hundreds of thousands of observations nested within hundreds of general practices. We have described how weighted regression and meta-analysis can be used for the purpose of analysing tall routine datasets. We have compared these methods to an existing subsampling approach[7], through application to electronic health records data from two contrasting examples, where regression coefficient estimates obtained through weighted regression and meta-analysis were similar to subsampling approaches and to the conventional approach of analysing the entire dataset.

Drovandi et al.[7] proposed the use of experimental design techniques to extract the information needed to answer the specific research question. This approach would be useful in methodological research, where it would be impractical to explore a number of different methods on a very large dataset. The disadvantage of subsampling is that it is difficult to know whether the data extracted are representative of the entire dataset. In our applications to case study examples, results based on an acquired subsample showed some discrepancies from the complete data analysis. We stopped the subsampling algorithm at reaching a pre-specified subsample size as in Drovandi et al., but other researchers could potentially obtain closer results to the complete data analysis by stopping the algorithm where there is little change in parameter estimates and corresponding standard errors with each additional iteration of the algorithm. Another possible explanation for discrepancies in results might be that the data were extracted based on covariate values alone and not the nesting variables (in our case only general practice). It would be possible to select data by nesting variables such as practice; in the subsampling algorithm one could choose data based on the practice-level effects and covariate values that give the maximum value of the utility function. Although this method gave improved results in our example applications, we do not report the results in this paper because the method was not computationally efficient; the number of observations per covariate combination and practice was very small relative to the size of our example datasets and so it took over 24 hours to proceed through the subsampling algorithm.

Another limitation of the subsampling method is that we consider all combinations of covariate values observed in the entire dataset for inclusion in our subsample. This would not be practical in other datasets with very many variables, since there are likely to be many possible covariate combinations that are available for selection from the dataset. In datasets where some values of covariates are not very frequent, the approach for extracting covariate combinations could be improved by extracting covariate combinations that occur within a specified window.[7,19] This will be most relevant for datasets with continuously measured variables.

Where it is preferable to use all of the data, we have found weighted regression and meta-analysis techniques to perform well. Both methods involve fitting the regression models to smaller datasets, thus reducing the time required for statistical analysis. In case study 1, the weighted regression approach reduced the size of the dataset substantially to 13% of its original size. Thus, this method facilitates analysis in software that currently cannot handle tall data, for example enabling Bayesian analyses in *WinBUGS*.[20] Although weighted regression is useful, there would be little gain in using this approach where the outcome of interest is very variable. Before attempting this approach we recommend checking that the number of possible outcome values is small relative to the size of the entire dataset, so that it is possible to collapse the dataset to a much smaller size. For similar reasons, weighted regression is limited in the presence of a continuously measured covariate. Our approach was to categorise any continuous covariate into fairly broad groups before collapsing the dataset to a smaller size, and to use the mean covariate value for each group in the regression analysis. This approach performed well in the applications to data from both case studies 1 and 2, but it does introduce measurement error and may yield larger standard errors for regression coefficients in other datasets. A further limitation of weighted regression is that it scales badly, becoming increasingly slow as the number of covariates increases.

In the presence of a continuous covariate, we have found meta-analysis techniques to be useful and we expect that this would also be the case where the outcome is continuously measured. For our purpose of fitting a regression model with a practice-level effect, meta-analysis approaches were fast to implement because the division of the data by practice removed the need for the practice-level effect in the analysis of data from each subgroup. We note that computing time could be reduced further by combining the weighted regression and meta-analysis approaches.

An important feature of the meta-analysis approach is that one may not be estimating the same quantity as a complete data analysis, especially for non-linear link functions. In particular, a single model would typically be specified so that the effects of the confounding variables are taken to be the same across all subgroups. In contrast, the meta-analysis approach allows confounder effects to vary across subgroups and, as such, provides a more flexible approach to the control of confounding. This is a reason for some differences between estimates obtained through conventional analysis of the entire dataset and meta-analysis of practice-level results. In future work we plan to consider methods to estimate a common treatment effect while allowing confounder effects to vary across subgroups.

Our case studies were carried out as pre-specified in the original study protocols. We focussed on fitting two-level models to data comprising observations (level-one units) nested in general practices (level-two units). With more than two levels, the meta-analysis approaches could be used to analyse the data. One approach could be to divide the data into level-two units (e.g. practices), and recombine in multiple stages (one stage per level), or to divide into top level units (e.g. regions). In our datasets, it was natural to divide the data by practice, but there may not always be an obvious variable on which to partition the data. The use of experimental designs could assist in this.[7,19]

A limitation of the meta-analysis approaches is that we could not handle cross-classified data which may arise in studies of hospital patient outcomes, if there is interest in allowing both for between-hospital variation and between-practice variation. Another disadvantage of the meta-analysis approaches is that it would be impossible to fit a regression model to data from practices that show no variation in the outcome of interest. In this case we could divide the data at a higher level, for example by region, then fit a mixed model to data from each practice and combine random practice-level effects and fixed covariate effects across regions using meta-analysis techniques.

Univariate meta-analysis would be expected to perform reasonably well when the covariates are centred and uncorrelated. In case study 1, the between-practice variance in intercept was higher in the univariate meta-analysis than in the conventional analysis of the entire data. For the purpose of examining association between some patient characteristics and an outcome of interest, estimation of the intercept and random-effects distribution is not important. Further investigation is needed to determine the use of meta-analysis for other purposes, such as prediction modelling.

In the analysis of longitudinal data it may be required to fit models with more than one random effect, for example at the patient and practice levels. The meta-analysis approach could be used for this purpose. Before attempting this analysis of routine data, it is important to think about the longitudinal value of the data which has not been collected for the purpose of research.[21] Developing methods for the analysis of tall longitudinal data should form the subject of future work.

The primary aim in each case study is to estimate the effect of independent variables on an outcome of interest in the population. For the purpose of population-based inference, marginal models fitted by generalized estimating equations would have been less sensitive to parametric assumptions than likelihood-based methods and would have been computationally more efficient.[22] The computationally efficient methods described in this paper would facilitate fitting marginal models in the same way they have assisted in fitting generalised linear mixed models to tall data. We chose not to focus on marginal models here. In the protocols for the case studies considered in this paper it was of interest to allow for further nesting of observations within regions as well as practices, and this is something that is less straightforward to account for in marginal models. Further, marginal models are less useful for prediction, which is of interest in many studies using health records data, since no distribution is specified for the outcome of interest.

Findings based on analysis of routine data which have not been collected for the purpose of research are likely to be affected by numerous biases. Routine data may provide incomplete information which would need explicit strategies such as multiple imputation and data linkage to deal with. There is a further danger of selection bias, for example due to disease and event definitions relying on code groups and algorithms that are typically less accurate than prospectively applied definitions. Adjusting for multiple biases in the analysis of routine data would greatly increase the complexity of the mixed model and computing time. The methods described in this paper could potentially assist in this; a representative subsample could be used for methodological development and meta-analysis might facilitate the application of bias-adjustment methods to the entire dataset.

In summary, we have identified scalable methods for the analysis of routine datasets comprising a very large number of observations. The existing subsampling approach allows us to extract the data required to answer the research question and has performed well in example applications. However, it may be preferable to make use of all the data. Where all data are discrete, weighted regression is equivalent to fitting the model to the entire dataset. In the presence of a continuously measured covariate, we have found meta-analysis techniques to be very useful. Both weighted regression and meta-analysis approaches are accessible to applied researchers and are easily implemented in standard statistical software.

**Acknowledgements**

The authors are grateful to the CPRD team at the University of Cambridge. In particular, we thank Carol Wilson and Anna Cassel for providing access to the case study datasets that they spent much time preparing for analysis. Kirsty Rhodes was funded by Medical Research Council Unit Programmes U105260558 and MC_UU_00002/5. Rebecca Turner and Ian White were funded by Medical Research Council Unit Programmes U105260558 and MC_UU_12023/21.

**Table 1** Case study 1: Mixed effects logistic regression to investigate the influence of QOF indicators on the first recording of elevated cholesterol in cases of severe mental illness

| Covariates | Entire dataset | | Subsampling approach | | Weighted regression* | | Univariate meta-analyses of practice data† | | Multivariate meta-analysis of practice data | |
|---|---|---|---|---|---|---|---|---|---|---|
| No. of observations | 2,116,948 | | 200,087 | | 273,685 | | 2,116,948 | | 2,116,948 | |
| | RD | SE | RD | SE | RD | SE | RD | SE | RD | SE |
| SMI x intervention1 | 0.04 | 0.02 | 0.07 | 0.04 | 0.04 | 0.02 | 0.03 | 0.05 | 0.03 | 0.03 |
| SMI x (year=2004) x intervention1 | 0.003 | 0.001 | -0.008 | 0.002 | 0.004 | 0.001 | 0.01 | 0.02 | 0.004 | 0.004 |
| SMI x (year=2010) x intervention1 | 0.002 | 0.001 | 0.003 | 0.001 | 0.002 | 0.001 | 0.02 | 0.03 | 0.004 | 0.004 |
| SMI x intervention2 | 0.09 | 0.04 | 0.14 | 0.05 | 0.10 | 0.05 | 0.05 | 0.08 | 0.08 | 0.07 |
| SMI x (year=2011) x intervention2 | 0.08 | 0.02 | 0.07 | 0.01 | 0.09 | 0.03 | 0.06 | 0.09 | 0.07 | 0.06 |
| SMI x (year=2014) x intervention2 | 0.008 | 0.002 | 0.02 | 0.004 | 0.009 | 0.003 | 0.02 | 0.03 | 0.009 | 0.009 |
| Between-practice variance $\tau^2$ | 0.05 | | 0.04 | | 0.05 | | 2.68 | | 0.05 | |
| Time to implement the method (in seconds)‡ | 38,721 (11 hours) | | 16,910 (4.7 hours) | | 285 (4.8 minutes) | | 68 (1.1 minutes) | | 101 (1.7 minutes) | |

RD risk difference; SE standard error; intervention1 denotes the first QOF indicator introduced in 2004; intervention2 denotes the second QOF indicator introduced in 2011. *We used the mean age for each group defined by values of outcome, covariate values and practice. †Each row corresponds to a separate univariate meta-analysis of regression coefficient estimates. ‡Average of three measurements of the time taken to implement the procedure using *R* (version 3.3.1)[9] on a computer with a 64-bit Windows operating system, Intel Core i7 processor, 3.4GHz speed and 16 GB installed memory (RAM).

**Table 2** Levels of each covariate available for selection in the subsampling approach

|  | Covariate | Levels |
|---|---|---|
| **Case study 1** | time<br>SMI<br>intervention 1<br>intervention 2<br>age (standardised)<br>gender | 1,2,3,4,5,…,19<br>0,1<br>0,1<br>0,1<br>-1,-0.5,0,0.5,1,1.5,2<br>0,1 |
| **Case study 2** | age (standardised)<br>gender<br>no. of morbidities: moderate 2-3<br>high 4-5<br>very high 6+ | -1.5,-1,-0.5,0,0.5,1,1.5,2,2.5<br>0,1<br>0,1<br>0,1<br>0,1 |

**Table 3** The first ten covariate combinations selected at each iteration of the existing subsampling approach.

|  | Covariate | Iteration of algorithm | | | | | | | | | |
|---|---|---|---|---|---|---|---|---|---|---|---|
|  |  | 1 | 2 | 3 | 4 | 5 | 6 | 7 | 8 | 9 | 10 |
| **Case study 1** | time | 19 | 17 | 16 | 8 | 10 | 1 | 17 | 19 | 17 | 19 |
|  | SMI | 1 | 1 | 1 | 1 | 1 | 1 | 1 | 0 | 0 | 1 |
|  | intervention 1 | 1 | 1 | 1 | 0 | 1 | 0 | 1 | 1 | 1 | 1 |
|  | intervention 2 | 1 | 1 | 0 | 0 | 0 | 0 | 1 | 1 | 1 | 1 |
|  | age (standardised) | 2 | 2 | 2 | 2 | 2 | 2 | -1 | 2 | 2 | -1 |
|  | gender | 0 | 0 | 0 | 0 | 0 | 0 | 1 | 1 | 1 | 0 |
|  | No. of observations | 514 | 753 | 826 | 1536 | 1363 | 1504 | 1853 | 1994 | 2197 | 1228 |
|  |  | 1 | 2 | 3 | 4 | 5 | 6 | 7 | 8 | 9 | 10 |
| **Case study 2** | age (standardised) | -1.5 | -1.5 | -1 | -1 | 2.5 | -0.5 | -1.5 | -1.5 | -0.5 | 2.5 |
|  | gender | 0 | 1 | 0 | 1 | 1 | 0 | 1 | 0 | 1 | 0 |
|  | no. of morbidities: moderate 2-3 | 0 | 0 | 0 | 0 | 0 | 0 | 0 | 0 | 0 | 0 |
|  | high 4-5 | 0 | 0 | 0 | 0 | 0 | 0 | 1 | 1 | 0 | 0 |
|  | very high 6+ | 1 | 1 | 1 | 1 | 1 | 1 | 0 | 0 | 1 | 1 |
|  | No. of observations | 2 | 5 | 28 | 17 | 162 | 90 | 34 | 74 | 40 | 467 |

**Table 4** Case study 2: Mixed effects negative binomial regression to investigate the association between multimorbidity and rates of primary care consultations.

| Covariates | Entire dataset | | Subsampling approach | | Weighted regression* | | Univariate meta-analyses of practice data†‡ | | Multivariate meta-analysis of practice data‡ | |
|---|---|---|---|---|---|---|---|---|---|---|
| No. of observations | 349,785 | | 36,150 | | 67,627 | | 349,785 | | 349,785 | |
| | RR | SE | RR | SE | RR | SE | RR | SE | RR | SE |
| Intercept | 18.70 | 0.40 | 18.62 | 0.61 | 18.70 | 0.40 | 18.73 | 0.56 | 18.73 | 0.56 |
| age (standardised) | 1.24 | <0.001 | 1.15 | 0.004 | 1.24 | <0.001 | 1.25 | 0.002 | 1.25 | 0.002 |
| gender (Male) | 0.68 | <0.001 | 0.87 | 0.008 | 0.68 | <0.001 | 0.68 | 0.002 | 0.68 | 0.002 |
| no. of morbidities (vs. low 0-1) | | | | | | | | | | |
| moderate 2-3 | 2.11 | <0.001 | 2.21 | 0.02 | 2.11 | <0.001 | 2.10 | 0.008 | 2.10 | 0.008 |
| high 4-5 | 2.95 | <0.001 | 2.95 | 0.04 | 2.95 | <0.001 | 2.92 | 0.02 | 2.92 | 0.02 |
| very high 6+ | 3.94 | 0.002 | 3.86 | 0.05 | 3.94 | 0.002 | 3.86 | 0.04 | 3.86 | 0.04 |
| Between-practice variance $\tau^2$ | 0.40 | | 0.34 | | 0.40 | | 0.40 | | 0.39 | |
| Dispersion parameter $\theta$ | 1.25 | | 1.75 | | 1.25 | | $1.30^4$ | | $1.30^4$ | |
| Time to implement the method (in seconds)§ | 5,959 (1.7 hours) | | 1,954 (33 minutes) | | 1,585 (26 minutes) | | 16 | | 27 | |

RR rate ratio; SE standard error. *We used the mean age for each group defined by values of outcome, covariate values, practice and the offset variable, time in the study. †Each row corresponds to a separate univariate meta-analysis of regression coefficient estimates. ‡Dispersion parameter estimates weighted according to the size of the practice and averaged across practices. §Average of three measurements of the time taken to implement the procedure using *R* (version 3.3.1)[9] on a computer with a 64-bit Windows operating system, Intel Core i7 processor, 3.4GHz speed and 16 GB installed memory (RAM).